\documentclass[referee]{aa}
\usepackage{graphicx}

\usepackage{longtable}
\usepackage{natbib}
\bibpunct{(}{)}{;}{a}{}{,}

\begin{document}
\title{Resetting chemical clocks of hot cores based on S--bearing molecules}

        \author{V. Wakelam\inst{1}, P. Caselli\inst{2},
        C. Ceccarelli\inst{3}, E. Herbst\inst{4} and A. Castets\inst{1}}
               \institute{
Observatoire de Bordeaux, BP 89, 33270 Floirac, France
\and INAF--Osservatorio Astrofisico di Arcetri, Largo E. Fermi 5, 50125
Firenze, Italy
\and Laboratoire d'Astrophysique, Observatoire de Grenoble - BP 53,
F-38041 Grenoble cedex 09, France
\and Departments of Physics, Chemistry, and Astronomy, The Ohio State
University, Columbus, OH 43210, USA}
        \offprints{wakelam@obs.u-bordeaux1.fr}

         \date{Received {\today} ; accepted}
\titlerunning{S-molecules as chemical clocks}
\authorrunning{Wakelam et al.}

         \abstract{
We report a theoretical study of sulphur chemistry, as
applied to hot cores, where S-bearing molecular
ratios have been previously proposed and used as chemical clocks.
As in previous models, we follow the S-bearing molecular composition after
the injection of  grain mantle components into the gas phase.
For this study, we developed a time-dependent chemical model
with  up-to-date reaction rate coefficients.
We ran several cases,
using different realistic chemical compositions for the grain mantles
and for the gas prior to mantle evaporation.
The modeling shows that  S-bearing molecular ratios
depend very critically on the gas temperature and density, the
abundance of atomic oxygen, and, most
importantly, on the form
of sulphur injected in the gas phase,  which is very poorly known.
Consequently, ratios of S-bearing molecules cannot be easily used
as chemical clocks. However, detailed observations and
careful modeling of both physical and chemical structure
can give hints on the source age and constrain the mantle composition
(i.e. the form of sulphur in cold molecular clouds) and,
thus, help to solve the mystery of the sulphur depletion.
We analyse in detail the cases of Orion and IRAS16293-2422. The
comparison of the available
observations with our model suggests that the majority of sulphur
released from the mantles is mainly in, or soon converted into,
atomic form.

}

\maketitle

\section{Introduction}

It is a long-standing dream to use relative abundances of different
molecules as chemical clocks to measure the ages of astronomical
objects.
Studies of the ages of star formation regions have recently focused
on S-bearing molecules.
\citet{1997ApJ...481..396C} and \citet{1998A&A...338..713H} were the first to
propose that the relative abundance ratios of SO, SO$_2$ and
H$_2$S could be used to estimate the age of
the hot cores of massive protostars.
The underlying idea is that the
main reservoir of sulphur is H$_2$S on grain mantles,
and that when the hot core forms, the
mantles evaporate, injecting the hydrogen sulphide into the gas phase.
Endothermic reactions in the hot gas convert H$_{2}$S into atomic
sulphur and SO from which more SO and, subsequently, SO$_{2}$ are
formed, making the SO$_{2}$/SO and SO/H$_2$S
ratios nice functions of time.
These studies have triggered a variety of work, both observational
and theoretical
\citep{1997ApJ...481..396C,
1998A&A...338..713H,2003A&A...399..567B}.

This line of research, however, has been challenged by recent ISO
observations, which have cast doubt on the basic assumption
      that sulphur is mainly trapped in grain mantles as H$_2$S. The
      lack of an appropriate feature in the ISO spectra of high
      \citep{2000ApJ...536..347G} and low \citep{2000A&A...360..683B}
      mass protostars sets an upper limit on the mantle H$_{2}$S abundance
      which cannot exceed about $10^{-7}$ with respect to H$_{2}$
      \citep{1998ARA&A..36..317V}.
        Indeed, the identity of the major
reservoir of
sulphur in cold molecular clouds is a long standing and unresolved problem,
for the sum of the detectable S-bearing molecules is only a very small
fraction of the elemental S abundance \citep{1994A&A...289..579T}.
Since sulphur is known not to be depleted in the diffuse medium
\citep[e.g.][]{1994ApJ...430..650S}, it is usually
assumed that sulphur in dense clouds is depleted onto the grain
mantles rather than in refractory cores
\citep*[e.g. ][]{1994ApJ...421..206C},   but how this happens is a
mystery. In a theoretical study, \citet{1999MNRAS.306..691R}
proposed that in collapsing translucent clouds sulphur
is efficiently adsorbed onto grain mantles.  In fact, in these regions, most of
the gas-phase sulphur is in the form of S$^{+}$, while grains are typically
negatively charged, so that the collisional cross section for sulphur is
enhanced compared with neutral species (e.g. O) and sulphur is removed
from the gas phase more rapidly.

Another mystery is the form of sulphur on dust grains.
The simplest possibility is that it consists of relatively isolated atoms, as would
occur in a matrix, or perhaps as isolated pairs of atoms (S$_{2}$).
Another possibility is that the sulphur is amorphous (or even
crystalline), having formed islands of material from the initially
adsorbed atoms.  Crystalline sulphur is known to come in two forms - 
rhombic and monoclinic - both of which consist of S$_{8}$ cyclic 
molecules.  Vaporization leads to a complex mixture of sulfur 
polymers through S$_{8}$ in complexity.
If sulphur is elemental and amorphous, evaporation
is also likely to lead to molecules of sulphur  through
eight atoms in complexity.
So far, the only S-bearing
species firmly detected on granular surfaces is OCS, but with
a relatively low fractional abundance of $10^{-7}$ \citep{1997ApJ...479..839P}.
Recently, \citet{2002Natur.417..148K} claimed the detection of iron sulphide
(FeS) grains in protoplanetary disks, but there is no evidence to suggest that
solid FeS is the main form of sulphur in the parent collapsing
environment. Actually, if the main form of solid sulphur is FeS,
S should follow Fe depletion, which is not observed
\citep{1994ApJ...430..650S}.
Even more recently, \citet{2003MNRAS.341..657S} suggested that
hydrated sulphuric acid
(H$_{2}$SO$_{4}$ $\cdot$ H$_{2}$O) is the main sulphur reservoir.
In whatever form sulphur resides in the grain mantles, there is the possibility
that the species, once evaporated, are very quickly destroyed to give
atomic sulphur.
In summary, although all the evidence is that sulphur is depleted
onto grain mantles in cold clouds, its particular form is very uncertain.

Given the need for  chemical clock methods,  it is timely to reconsider
  the
use of S-bearing molecules in this fashion.
In this paper, we present a model with an up-to-date chemical
network involving S-bearing molecules.
We run several cases to cover a large, realistic parameter
space for hot core sources, consistent with
present observational constraints.
Based on the results we obtain, we conclude that
it is tricky to use abundance ratios of S-bearing
molecules as chemical clocks in the absence of other
constraints,
for they depend more on the initial
conditions, gas density, temperature, and the initial form of sulphur
injected in the gas phase than on the age of the
source.

The paper is organized as follows:
we describe the model in \S 2, the model results in \S 3,
and in \S 4 we discuss the practical consequences of those results
and apply the model to the specific cases of Orion and IRAS16293-2422.

\section{The Model}

We have developed a
pseudo-time dependent model for the gas phase chemistry
that computes the evolution of the chemical composition of
a volume of gas with a fixed density and  temperature.
Our goal is to follow how the S-bearing molecular abundances
vary with time when the gas undergoes a sudden change in its
temperature and density, and/or in its overall chemical abundance,
because of the evaporation of grain mantles.
In hot cores the dust temperature increases to an extent that it
exceeds the mantle evaporation
temperature, i.e. $\sim$100 K, and all the components of the grain mantles
are suddenly injected into the gas phase, similarly to what has been done
in previous studies of hot cores 
\citep[e.g.][]{1988MNRAS.231..409B,1992ApJ...399L..71C,1993ApJ...408..548C,
1997A&A...325.1163M}. In fact, it is more probable that hot cores have
spatial gradients in temperature and density releasing the molecules
at different
times depending on their surface binding energies
\citep[see 
][]{1996ApJ...471..400C,1999MNRAS.305..755V,2002A&A...389..446D,2003ApJ...585..355R}.
However, the goal of this work is mainly to test
the effects of the form of the main initial sulphur bearing molecules and we
preferred to simplify the problem assuming that all the molecules
evaporate simultaneously from the grain. More detailed models will be
presented in a forthcoming paper.

In order to simulate these conditions, the  gas-phase
chemical composition prior to evaporation of the
mantles
     is taken to be similar to that of dark molecular clouds.
At time $t$= 0, the grain mantle components are injected into the gas phase,
  and the model follows the changes
in the gas chemical composition with a given gas temperature and 
density.  Throughout this paper we will
use the word ``evaporation'' to refer to the loss of the grain
mantles.

The model is a reduced chemical network, which includes 930 reactions
involving 77 species containing the elements H, He, C, O and S.
The standard neutral-neutral and ion-neutral reactions are considered.
Most of the reaction coefficients are from the NSM (``new standard
model''; {\it 
http:$//www.physics.ohio-state.edu/~eric/research_{-}files/cddata.july03$})
database; see also \citet*{1996A&AS..119..111L}, updated with new
values or new analyses of assorted values in databases (e.g. the NIST 
chemical kinetics database at  {\it 
http://kinetics.nist.gov/index.php}) when
available. Furthermore, several high temperature (neutral-neutral)
reactions have been added.
To select the reduced network, we have followed
\citet{2002A&A...381L..13R} for CO formation,
\citet{1979ApJS...41..555H} and \citet{1980ApJ...236..182H} for the
oxygen chemistry and \citet{1993MNRAS.262..915P} and
\citet{1997ApJ...481..396C} for the sulphur chemistry.

To validate this network, we compared our
results with abundances previously obtained by
\citet*{1996A&AS..119..111L} at low temperature and
\citet{1997ApJ...481..396C} at higher
temperatures using the same initial abundances as these authors. We
found that we can reproduce molecular abundances to better than a factor of
three. This is an indication that small variations in the rate
coefficients between the updated NSM and UMIST databases do not 
strongly influence the computed abundances of sulphur bearing 
species.
One exception is the CS molecule, which we produce at an abundance ten
times less than Charnley's model, because our adopted rate coefficient for the
reaction CO + CRPHOT $\rightarrow $ O + C, (where CRPHOT is a photon
induced by cosmic rays) is 50 times smaller
than in the UMIST database. The lowered abundance of C then
translates into a lowered abundance for CS, since C is a precursor of
CS.

Note that we have assumed  the gas to be totally shielded from the
interstellar UV field
and no other UV field to be present.
Thus, the model does not include any photochemistry, with the
exception of cosmic ray--induced photodestruction reactions.
The model takes into account a reduced gas-grain chemistry:
H$_2$ is formed on grain surfaces and the recombination of ions with
negatively charged grains occurs \citep[see ][ for the recombination of ions with negative grains]{1999ApJ...527..262A}.
Moreover, neutral species can deplete onto grain mantles, and
mantle molecules can evaporate because of thermal effects and cosmic rays
\citep*{1992ApJS...82..167H,1993MNRAS.261...83H}.

\subsection{Gas-phase and mantle abundances prior to evaporation}

To help determine a set of molecular abundances prior to mantle evaporation,
we  ran the model with a temperature equal to 10 K and a density
$\rm n_{H_{2}}$ equal to $10^4$ cm$^{-3}$,  including freeze out, for 10$^7$
yr.  At this time,  species such as SO, SO$_2$, and
CS reach abundances similar to those observed in dark clouds \citep[$\sim$
10$^{-9}$, $\leq 10^{-9}$, and $\sim 10^{-9}$, respectively;
][]{2000ApJ...542..870D}.
The adopted
elemental abundances (with respect to H$_2$) for He, O, C$^{+}$ and
S$^{+}$ are respectively:
0.28, $6.38 \times 10^{-4}$ \citep{1998ApJ...493..222M}, $2.8 \times
10^{-4}$ \citep{1996ApJ...467..334C} and $0.3 \times 10^{-4}$; the 
sulphur abundance refers to the gaseous and grain
portions (see below).
The late-time abundances obtained could
not  be reasonably used without modification for the pre-evaporated
chemical composition for several reasons.  First, our model, like
most other gas-phase treatments
\citep[e.g.][]{1996A&AS..119..111L,1997A&AS..121..139M},
overestimates the O$_{2}$ and H$_{2}$O abundances in cold dense
clouds by orders of magnitude
with respect to the ISO, SWAS and ODIN observations
\citep[e.g.][]{2003ApJ...582..830B,2003A&A...402L..77P}.  Secondly,
the elemental sulphur abundance pertaining to the gas must be
lowered to avoid getting very
high abundances of sulphur-bearing species. The portion of the
abundance not used for the gas can be considered to reside in grain mantles until evaporation or in grain core.
In order to mimic realistic conditions, we thus adopted three different
      compositions for the pre-evaporative gas, as follows:

{\it Composition A}:
We adopted the computed late-time molecular abundances except for
O$_{2}$ and H$_{2}$O, which were assumed to be $10^{-7}$ and
$10^{-8}$ with respect to H$_{2}$
respectively, in agreement with
observations in molecular clouds , and for atomic
oxygen, O, which was assumed to carry
the oxygen not locked into CO, leading to a fractional abundance of
$2.6 \times 10^{-4}$, as suggested
by observations \citep[e.g. ][]{1997A&A...322L..33B,1999A&A...347L...1C,
2002ApJ...581..315V,2001ApJ...561..823L}. In addition, the initial
gas-phase sulphur abundance was taken to be
a factor of 30 lower than the elemental abundance.

{\it Composition B}:
We re-computed the late time abundances, lowering artificially by two
orders of magnitude the rate of the dissociative recombination of
H$_{3}$O$^{+}$, to decrease the computed O$_{2}$ and H$_{2}$O abundances.
In this case, it was only necessary to lower the initial elemental sulphur
abundance by a factor of five. The abundance of atomic oxygen in this
case is $4.8\times 10^{-4}$, consistent with observations in molecular clouds.

{\it Composition C}:
The abundances were taken to equal  those
measured in the direction of L134N, as reported in  Table 4 of
\citet*{2001A&A...378.1024C}. The oxygen not contained in the
species reported in Table 4 was assumed to be in atomic form.

All three gas-phase compositions have large abundances of
atomic oxygen,  in agreement with observations. We assume that this large
O abundance is also present at the beginning of the hot core phase,
in contrast with previous studies.  This implies many
differences in the computed abundances, as shown
in Section 3.1 and discussed in Section 4.1.
Table \ref{ini-abu} lists the abundances of the main S-bearing
molecules for the three gas-phase compositions  prior to  mantle
evaporation.

\begin{table}
\caption{Computed late time (A, B) and adopted (C) gas-phase abundances
with respect to H$_2$ for the
main S-bearing
molecules prior to mantle evaporation.\label{ini-abu}}

         \begin{tabular}{l|c|c|c}
\hline
Species & A & B & C \\
\hline
SO     & $9\times 10^{-9}$  & $3\times 10^{-9}$ & $3.1\times 10^{-9}$\\
SO$_2$ & $4\times 10^{-10}$ &$1.5\times 10^{-10}$ & $1\times
10^{-9}$$^{a}$\\
H$_2$S & $1\times 10^{-10}$ & $1\times 10^{-11}$ & $8\times 10^{-10}$\\
CS     & $4\times 10^{-10}$ & $2\times 10^{-8}$ & $1.7\times 10^{-9}$\\
\hline
\end{tabular}
\small{
\begin{list}{}{}
\item[$^{a}$] Here we took the value of the upper limit.
\end{list}
}
\end{table}

At time $t$ = 0, the grain mantle components are injected into
the gas phase.
The abundances of the major mantle components  are relatively well
constrained by the observations and we took the observed abundances
(with respect to H$_2$) in
high mass protostars: H$_{2}$O: $10^{-4}$
\citep{1996A&A...315L.333S}, H$_{2}$CO: $4 \times 10^{-6}$
\citep{2001A&A...376..254K}, CH$_{3}$OH: $4 \times 10^{-6}$
\citep{1996ApJ...472..665C} and CH$_{4}$: $10^{-6}$
\citep{1998A&A...336..352B}. Note that in order to shorten the number of treated species, we negleted the CO$_2$ molecule which is abundant in mantles \citep[$2\times 10^{-5}$ with respect to H$_2$,][]{1999ApJ...522..357G}, because it is not a crucial element of the sulphur chemistry. On the contrary, the situation
is very uncertain with respect to the S-bearing mantle molecules, as
discussed in the Introduction.
In order to study the influence of the injected S-bearing abundances
on the evolution of the chemical composition, we have run models
with four types of material evaporating from mantles
that differ in their major
sulphur-bearing species, as reported in Table \ref{iniabundS}.
In practice, sulphur on the grain mantles can be stored as
OCS, H$_{2}$S,  or pure sulphur in a matrix-like, amorphous, or
even crystalline form.  Of the four mixtures, the first one
(used in Model 1) has the bulk of the sulphur in the refractory core of the
grain, while the other three (used in Models 2-4) have large abundances of
elemental sulphur leading upon evaporation directly or eventually
to either S
or S$_2$ in the gas.

\begin{table}
         \caption{Adopted abundances for mixtures of evaporated S-bearing
molecules (with respect to H$_{2}$).\label{iniabundS}}

\begin{tabular}{l|cccc}
\hline
Species & Mod. 1 & Mod. 2 & Mod. 3 & Mod. 4 \\
\hline
OCS & $10 ^{-7}$ & $10 ^{-7}$ & $10 ^{-7}$ & $10 ^{-7}$ \\
H$_{2}$S & $10^{-7}$ & $10^{-7}$ & $10^{-7}$ & $10^{-8}$ \\
S & 0 & $3 \times 10^{-5}$ & 0 & $3 \times 10^{-5}$ \\
S$_{2}$ & 0 & 0 & $1.5 \times 10^{-5}$ & 0 \\
\hline
\end{tabular}
\end{table}

\subsection{"Important" reactions}

As the post-evaporative gas-phase chemistry proceeds, it is important to
determine the reactions that will {\it influence} the formation and
destruction of sulphur-bearing species the most severely.
We will call them "important" reactions in the following discussion.
By this term, we mean quantitatively those reactions that lead to 
significant variations of
the main S-bearing abundances when the relevant reaction rate is changed by
a small amount (specifically 10\%).
  Although such a determination has not
been featured in papers on astrochemistry, we thought it worthwhile to
introduce a suitable procedure here.
The aim of this study is twofold: (i)  to determine a set of 
reactions to check carefully in laboratory
experiments because the computed abundances are particularly sensitive to those
reactions, and (ii)  to ascertain whether different chemical networks
will lead to different results, and why.
To identify these reactions, we first defined a ``perturbation'' in 
the rate coefficient for
each of the 750 gas--phase reactions by multiplying them by a factor 
of 1.1, one at a time. 
  For each single perturbation, we then
computed the  abundance ``variations'' $\Delta X$ by
comparing the computed abundances after $10^4$~yr with the reference
abundances calculated with
the non-modified set of reaction rates, according to the expression
 $\Delta X = \frac{|X_{ref}-X|}{X_{ref}}$,
where $X_{ref}$ is the reference abundance and $X$ the abundance
obtained with the perturbed rate.
\begin{figure}[h]
\begin{center}
\includegraphics[width=\linewidth]{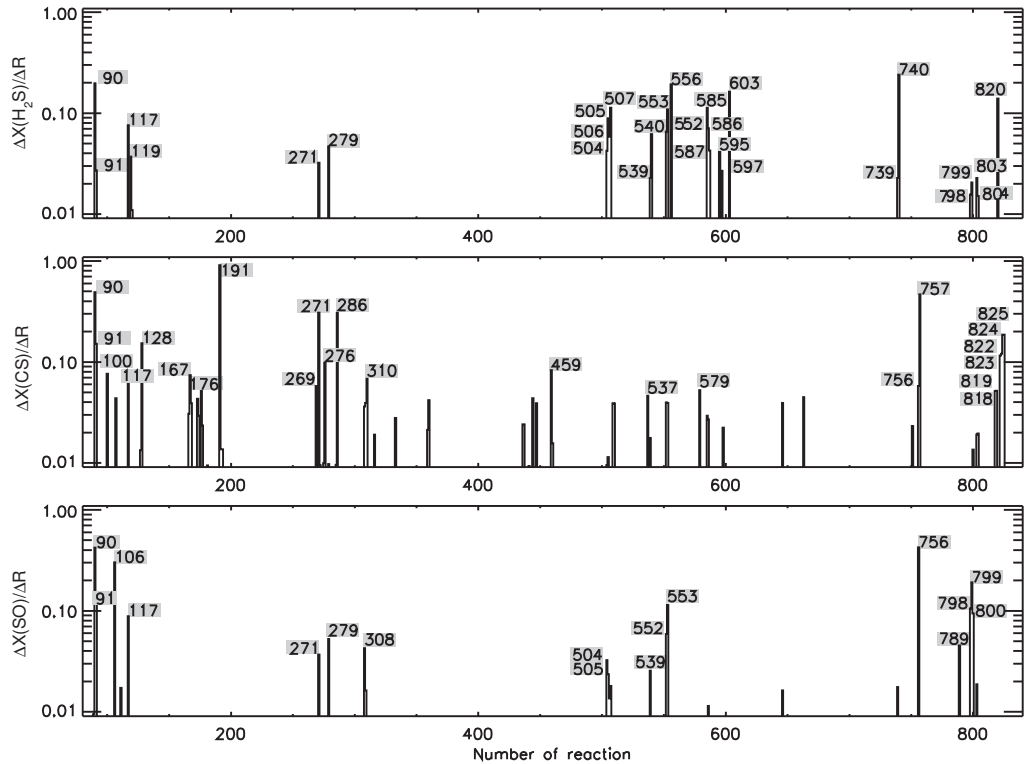}
\caption{The ``variation'' $\Delta X$ (see text) of the abundances
of SO, H$_2$S, and
CS divided by the  amplitude ($\Delta$R = 0.1) of the perturbation of the rate
coefficients plotted for reactions with number between 86 and 834;
  i.e., all the reactions in the model exept those
dealing with dust grains.
  The numbers
indicate those reactions that induce
a variation of more than 5\% of the highest variation for each
species. }
\label{3var}
\end{center}
\end{figure}

Fig~\ref{3var} shows the variations of the abundances of the
main S-bearing species H$_2$S, CS, and SO divided by the amplitude of
the perturbation ($\Delta$R = 0.1) for a temperature of 100~K and a 
density of $10^7$~cm$^{-3}$. The calculation has been performed for 
composition A and Model 2. The abscissa consists of the numbers of 
reactions in our network from 86 to 834. Vertical lines are included 
for those reactions that produce a normalized variation larger than 
1\% of the largest variation, while the actual numbers of reactions 
with a normalized variation greater than 5\% are listed.  Note that 
for a variation (0.1) equal in size to the amplitude of the 
perturbation, the line extends upward to unity and that a $\Delta X/\Delta R$ of 0.5 implies that the abundance of the studied molecule change by 5\% in abundance upon a 10\% change in the reaction rate.
The sets of numbered ``important'' 
reactions for the molecules H$_{2}$S, CS, and SO consist of 28, 24, 
and 17 reactions, respectively.
We do not show the variations of OCS and SO$_2$ because  the OCS
abundance is only affected significantly
(9\%) if reaction 757 is perturbed,  while SO$_2$ shows the
same behavior as SO so the corresponding figure is the same.

In Table~\ref{rates}, we list the 70 most important reactions for the 
chemistry of H$_2$S, CS,  SO, SO$_{2}$ and OCS for initial 
composition A and the four models (see previous section)
at a temperature of 100~K and a density of $10^7$~cm$^{-3}$.
The first two columns
give the number of the reaction and the actual reaction, while the 
third column gives the  reaction rate coefficient $k$ in terms of the 
standard parameters
$\alpha $ (cm$^{3}$ s$^{-1}$), $\beta $ and $\gamma $ (K):
\begin{equation}
k =  \alpha (T/300)^{\beta } exp(-\gamma /T)
\end{equation}
where T is  the gas temperature. In this table, we report only
the rate coefficients that
differ by more than 10\% from the UMIST ones. In the last four
columns, the symbol $\bullet$ indicates that the reaction is
important for sulphur chemistry in the model of the corresponding
column.
The list of reactions in Table \ref{rates} depends only weakly on the 
choice of initial
composition mentioned in section~2.1, because the three initial
compositions A, B and C give
sufficiently similar abundances at $10^4$~yr.

Of course, for very different initial compositions, the list of 
important reactions could be slightly different, especially if the 
network is different. This is for example the
case of the \citet{1997ApJ...481..396C} model, where all reactions 
involving atomic oxygen are
not important, for no initial gaseous atomic oxygen is assumed to be 
present, whereas ion--molecule reactions involving molecular ions such as H$_3^+$, H$_3$O$^+$, and H$_3$CO$^+$, are crucial.

We found that the relative importance of the reactions in Table 
\ref{rates} depends more on the initial mantle composition (Model 1 
to 4) than on the pre-evaporation composition of the gas (A, B or C).
 For Model 1, the chemical network is simpler (i.e. fewer important reactions)
 than for Models 2, 3, and 4 because there is no initial S or S$_2$, so
 that many reactions forming CS as well as reactions with S and S$^+$ lose importance.
Models 2 and 3 have a similar list of important reactions but some
reactions with SO gain in importance for
Model 3. Models 2 and 4 are even more similar in their lists of
important reactions. A few reactions with H$_2$S become
less important for Model 4 because the initial abundance of H$_2$S in this
model is ten times less
than in Model 2 and only one reaction (number 550) becomes important.

\begin{longtable}
	{l|l|rrr|cccc}
\caption{List of the reactions important for the chemistry of the SO,
SO$_2$, OCS, H$_2$S and CS molecules for a temperature of 100~K and a 
density of $10^7$~cm$^{-3}$ with initial
composition A.\label{rates}}\\
\hline
Number & Reaction\footnote{Only rate coefficients that differ by more than 10
percent from the UMIST value are included.} & $\alpha $ &  $\beta $ &  $\gamma $ & Mod. 1 &
Mod. 2 & Mod. 3 & Mod. 4 \\
\hline
\hline
89 & H$_2$ +     CRP $\rightarrow $      H$^{+}$ + H + e$^{-}$ &
2.860e-19  & 0.00 & 0.00 & & & $\bullet$ & \\

90 &  H$_2$    + CRP    $\rightarrow $ H$_{2}^{+}$   + e$^{-}$  &  &
& & $\bullet$   &$\bullet$  & $\bullet$ & $\bullet$ \\

91 &  He    + CRP    $\rightarrow $ He$^{+}$   + e$^{-}$        & & &
& $\bullet$  &$\bullet$ &  $\bullet$  & $\bullet$ \\

100 & CH$_4$   + CRPHOT $\rightarrow $ CH$_2$   + H$_2$  & 6.08e-14&
0.00&       0.00  & & $\bullet$ & $\bullet$ & $\bullet$  \\

106 & H$_2$O   + CRPHOT $\rightarrow $ OH    + H   & 2.52e-14&  0.00&
0.00    & $\bullet$  & $\bullet$ & $\bullet$ & $\bullet$     \\

117 & S     + CRPHOT $\rightarrow $ S$^{+}$    + e$^{-}$    &
2.49e-14&       0.00&       0.00  & & $\bullet$ &   & $\bullet$    \\

119 & H$_2$S   + CRPHOT $\rightarrow $ S     + H$_2$   & 1.34e-13&
0.00&       0.00   &$\bullet$   & $\bullet$ & $\bullet$ &  \\

120 & H$_2$S + CRPHOT $\rightarrow $ H$_2$S$^{+}$ +   e$^{-}$ &
4.41e-14 & 0.00 & 0.00 &$\bullet$   & & $\bullet$ & \\

125 & SO  +    CRPHOT $\rightarrow $  SO$^{+}$   +  e$^{-}$ &
1.30e-14 & 0.00 & 0.00  & & & $\bullet$ & \\

127 & OCS + CRPHOT  $\rightarrow $ S + CO & 1.39e-13 & 0.00 & 0.00 &
$\bullet$ & & &\\

128 & OCS   + CRPHOT $\rightarrow $ OCS$^{+}$  + e$^{-}$    &
3.76e-14& 0.00& 0.00  & $\bullet$ & $\bullet$ &  $\bullet$ &
$\bullet$\\

167 & C     + SO     $\rightarrow $ CS    + O     & 3.50e-11&
0.00&       0.00     & & $\bullet$ & $\bullet$ & $\bullet$  \\

168 & C + SO$_2$ $\rightarrow $ CO  + SO & & & & & & $\bullet$ & \\

176 & CH$_2$   + S      $\rightarrow $ CS    + H$_2$ &   &  & &
&$\bullet$  & & $\bullet$  \\

191 & O     + CS     $\rightarrow $ CO    + S         & 1.94e-11&
0.00& 231.00   & $\bullet$ & $\bullet$ & $\bullet$ & $\bullet$  \\

207 & O$_2$ + S $\rightarrow $ SO + O & 2.30e-12 & 0.00 & 0.00 &
$\bullet$ & & &\\

269 & H$_{3}^{+}$   + O      $\rightarrow $ OH$^{+}$   + H$_2$   & &
&  & $\bullet$   &$\bullet$  &  $\bullet$ & $\bullet$  \\

271 & H$_{3}^{+}$ + H$_2$O $\rightarrow $ H$_3$O$^{+}$ + H$_2$ &
4.50e-09& -0.50&  0.00 &$\bullet$&$\bullet$& $\bullet$ & $\bullet$ \\

276 & H$_{3}^{+}$ + CH$_4$O $\rightarrow $ CH$_{3}^{+}$  + H$_2$O +
H$_2$ & 1.80e-09&  -0.50& 0.00 & & $\bullet$ & $\bullet$ & $\bullet$
\\

279 & H$_{3}^{+}$   + S      $\rightarrow $ HS$^{+}$   + H$_2$     &
&  & & &$\bullet$  & $\bullet$ & $\bullet$  \\

285 & H$_{3}^{+}$ +    SO $\rightarrow $   HSO$^{+}$ + H$_2$ &
8.40e-09 & -0.50 & 0.00 & & & $\bullet$ & \\

286 & H$_{3}^{+}$   + OCS    $\rightarrow $ HOCS$^{+}$ + H$_2$  &
3.80e-09&     -0.50& 0.00   &$\bullet$   & $\bullet$ & $\bullet$ &
$\bullet$  \\

308 & He$^{+}$   + H$_2$O    $\rightarrow $ OH    + He     + H$^{+}$
& 1.32e-09&     -0.50&       0.00 & & $\bullet$ &  $\bullet$ &
$\bullet$   \\

310 & He$^{+}$   + CO     $\rightarrow $ O     + C$^{+}$     + He &
5.50e-10&     -0.50&       0.00 & & $\bullet$ & $\bullet$ & $\bullet$
\\

316 & He$^{+}$ + CH$_4$O $\rightarrow $ CH$_{3}^{+}$ + OH + He  &
1.70e-09  &   -0.50   &   0.00 & & & $\bullet$ &\\

333 & He$^{+}$ + OCS $\rightarrow $ CS + O$^{+}$ + He & 8.40e-10 &
-0.50 & 0.00  & $\bullet$ & & &\\

436 & CH$_{3}^{+}$ + O $\rightarrow $ HCO$^{+}$ +   H$_2$   &
2.05e-10  &    0.00  &    0.00 & & & $\bullet$ &\\

437 & CH$_{3}^{+}$ + O $\rightarrow $ HOC$^{+}$ + H$_2$ & & & & & &
$\bullet$ &\\

447 & CH$_{3}^{+}$ + SO $\rightarrow $ HOCS$^{+}$ +  H$_2$  &
4.20e-09  &   -0.50   &   0.00 & & & $\bullet$ &\\

459 & CH$_4$   + S$^{+}$     $\rightarrow $ H$_3$CS$^{+}$ + H    &
1.40e-10&       0.00&       0.00 & & $\bullet$ & $\bullet$ &
$\bullet$    \\

504 & O     + HS$^{+}$    $\rightarrow $ S$^{+}$    + OH         & &
& &   &$\bullet$  & $\bullet$ & $\bullet$   \\

505 & O     + HS$^{+}$    $\rightarrow $ SO$^{+}$   + H            &
&  & &   &$\bullet$  & $\bullet$ & $\bullet$   \\

506 & O     + H$_2$S$^{+}$   $\rightarrow $ HS$^{+}$   + OH         &
&  &  &  &$\bullet$  & $\bullet$  & $\bullet$  \\

507 & O     + H$_2$S$^{+}$   $\rightarrow $ SO$^{+}$   + H$_2$
& &   & &  &$\bullet$  & $\bullet$  & $\bullet$  \\

537 & H$_2$O   + HCO$^{+}$   $\rightarrow $ CO    + H$_3$O$^{+}$ &
2.10e-09& -0.50&  0.00 & $\bullet$ & $\bullet$ & & $\bullet$    \\

539 & H$_2$O   + HS$^{+}$    $\rightarrow $ S     + H$_3$O$^{+}$
&  & & &  &$\bullet$  & $\bullet$  & $\bullet$   \\

540 & H$_2$O   + H$_2$S$^{+}$   $\rightarrow $ HS    + H$_3$O$^{+}$
& 7.00e-10&       0.00&       0.00 & & $\bullet$ &  $\bullet$ &
$\bullet$ \\

550 & H$_2$O$^+$ + H$_2$S  $\rightarrow $ H$_3$S$^+$ + OH & 7.00e-10
&    0.00  &    0.00 & & & & $\bullet$ \\

552 & H$_3$O$^{+}$  + H$_2$CO   $\rightarrow $ H$_3$CO$^{+}$ + H$_2$O
& 2.60e-09& -0.50& 0.00 & $\bullet$ & $\bullet$ & $\bullet$ &
$\bullet$ \\

553 & H$_3$O$^{+}$  + CH$_4$O   $\rightarrow $ CH$_5$O$^{+}$ + H$_2$O
& &  & &  $\bullet$  &$\bullet$  & $\bullet$ & $\bullet$   \\

556 & H$_3$O$^{+}$  + H$_2$S    $\rightarrow $ H$_3$S$^{+}$  + H$_2$O
& &  &  & $\bullet$  &$\bullet$  & $\bullet$ & $\bullet$   \\

579 & HCO$^{+}$  + OCS    $\rightarrow $ HOCS$^{+}$ + CO       &
1.50e-09&     -0.50& 0.00& $\bullet$ & $\bullet$ &   & $\bullet$
\\

585 & H$_2$CO  + S$^{+}$     $\rightarrow $ H$_2$S$^{+}$  + CO    &
1.10e-09&     -0.50&       0.00  & & $\bullet$ & $\bullet$ &
$\bullet$ \\

586 & H$_2$CO  + S$^{+}$     $\rightarrow $ HCO$^{+}$  + HS      &
1.10e-09&     -0.50&       0.00   & & $\bullet$ & $\bullet$ &
$\bullet$  \\

587 & H$_2$CO  + H$_3$S$^{+}$   $\rightarrow $ H$_3$CO$^{+}$ + H$_2$S&
&  & &  $\bullet$  &$\bullet$  & $\bullet$ & \\
595 & S     + H$_3$S$^{+}$   $\rightarrow $ H$_2$S$_{2}^{+}$ + H & &
& &   &$\bullet$  & $\bullet$ & \\

597 & S$^{+}$    + H$_2$S    $\rightarrow $ S$_{2}^{+}$   + H$_2$   &
6.40e-10&     -0.50&       0.00   & & $\bullet$ & $\bullet$ &\\

603 & H$_2$S    + SO$^{+}$    $\rightarrow $ S$_{2}^{+}$   + H$_{2}$O
& 1.10e-09&       0.00&  0.00 & & $\bullet$ &  $\bullet$ & $\bullet$
\\

614 & H$^{+}$ + H$_2$O $\rightarrow $ H$_2$O$^{+}$ +   H  & 7.30e-09
&  -0.50 & 0.00 & & & $\bullet$ &\\

624 & H$^{+}$ + SO $\rightarrow $ SO$^{+}$ +    H  &  1.40e-08 &
-0.50 &  0.00 & & & $\bullet$ &\\

739 & S     + HS$^{+}$    $\rightarrow $ HS    + S$^{+}$       &    &
& &   &$\bullet$  &    & $\bullet$  \\

740 & S     + H$_2$S$^{+}$   $\rightarrow $ H$_2$S   + S$^{+}$ & &  &
&   &$\bullet$  & $\bullet$ & $\bullet$  \\

751 & H$_2$ + CH$_{3}^{+}$ $\rightarrow $ CH$_{5}^{+}$ +   PHOTON & &
& & & & $\bullet$ &\\

756 & O     + SO     $\rightarrow $ SO$_2$   + PHOTON     & 3.20e-16&
-1.60&   0.00 &$\bullet$  & $\bullet$ & $\bullet$ & $\bullet$     \\

757 & CO    + S      $\rightarrow $ OCS   + PHOTON             &  &
& & $\bullet$   &$\bullet$  & $\bullet$ & $\bullet$  \\

789 & H$_3$O$^{+}$  + e$^{-}$     $\rightarrow $ OH    + H      + H
& &  & &  &$\bullet$  & & $\bullet$ \\

798 & H$_3$CO$^{+}$ + e$^{-}$     $\rightarrow $ CO    + H      +
H$_2$& &  &  &  &$\bullet$  & $\bullet$ & $\bullet$  \\

799 & H$_3$CO$^{+}$ + e$^{-}$     $\rightarrow $ HCO   + H      + H
& &  & & &  $\bullet$  & $\bullet$ & $\bullet$  \\

800 & H$_3$CO$^{+}$ + e$^{-}$     $\rightarrow $ H$_2$CO  + H
& &  & & &  $\bullet$  & & $\bullet$ \\

803 & CH$_5$O$^{+}$ + e$^{-}$     $\rightarrow $ CH$_4$O  + H
& &  & & &  $\bullet$  & $\bullet$  &\\

804 & CH$_5$O$^{+}$ + e$^{-}$     $\rightarrow $ H$_2$CO  + H$_2$
+ H & &  & & &  $\bullet$  & $\bullet$ &\\

812 & H$_3$S$^{+}$ + e$^{-}$ $\rightarrow $ HS + H + H & 1.00e-07 &
-0.50 & 0.00 & $\bullet$ & & &\\

818 & H$_3$CS$^{+}$ + e$^{-}$     $\rightarrow $ CS    + H$_2$     +
H   &  &  & & &  $\bullet$  & & $\bullet$ \\

819 & H$_3$CS$^{+}$ + e$^{-}$     $\rightarrow $ H$_2$CS  + H & & & &
&$\bullet$  & & $\bullet$  \\

820 & SO$^{+}$   + e$^{-}$     $\rightarrow $ S     + O             &
& & &  &$\bullet$  &  $\bullet$ & $\bullet$  \\

821 & HSO$^{+}$ +   e$^{-}$ $\rightarrow $ SO  +    H  & & & & & &
$\bullet$ &\\

822 & OCS$^{+}$  + e$^{-}$     $\rightarrow $ CO    + S   & 3.00e-07&
0.00&       0.00  & & $\bullet$ & $\bullet$ & $\bullet$    \\

823 & OCS$^{+}$  + e$^{-}$     $\rightarrow $ CS    + O     &   &  &
& &  $\bullet$  & $\bullet$ & $\bullet$   \\

824 & HOCS$^{+}$ + e$^{-}$     $\rightarrow $ CS    + OH   &  &  & &
$\bullet$   &$\bullet$  &   $\bullet$ & $\bullet$   \\

825 & HOCS$^{+}$ + e$^{-}$     $\rightarrow $ OCS   + H    &  &  & &
$\bullet$   &$\bullet$  & $\bullet$  & $\bullet$    \\

\hline
\end{longtable}

The above analysis of ``important" reactions refers to only one 
perturbation amplitude.
In general, since the equations are not linear, the amplitude of the 
perturbation may
influence the results in a non-linear way.  To check for non-linearity,
we also ran the case where each reaction rate is twice as large as
the "standard" one (i.e. a pertubation amplitude of 1.0, which doubles the rate of reaction),
and still obtained linear variations, so that normalized variations 
are independent
of amplitude.
One exception concerns CS, for which several reactions become
important in the latter case.
These reactions, listed in Table~\ref{perturbation2} for composition 
A and Model 2,
must be added to the ones shown in Fig.~\ref{3var} for CS.

Because rate coefficients are often dependent on the temperature, a 
change in this parameter can affect which reactions are important.  In
particular, an increase of
the temperature to 200~K makes some of the neutral-neutral reactions 
more important (Table~\ref{rates200K}) and some reactions (579, 789, 
798, 800, 804 and 825) of Table~\ref{rates} negligible. On the 
contrary, an increase of the H$_2$ density to $10^8$~cm$^{-3}$ does 
not change
  the results significantly.

Finally, from Tables~\ref{rates} and \ref{rates200K} and
Fig.~\ref{3var}, we can determine the reactions producing the largest
variations for the S-bearing species under a set of relevant 
conditions and a reasonable time (10$^{4}$ yr) for our models: 
reactions 90, 153,
190, 191, 740, 756 and 757.  Reaction 90 is the cosmic ray ionization 
rate of H$_{2}$, which is obviously important for starting the 
ion-molecule chemistry. Reactions 153 and 190 are neutral-neutral 
reactions important only at the higher temperature considered (200 
K).  The other reactions (cf Table~\ref{rates}) are a collection of 
ion-molecule (740), neutral-neutral (191), and radiative association
(756 and 757) processes.  The importance of these
reactions would probably  have been overlooked had this analysis not 
been done. It will  therefore be crucial to know the
rate coefficients  of the mentioned reactions with
high precision. Of these reactions, only the rate coefficient
for 153 is well determined in the laboratory, although the lowest
measured temperature (300 K) means that the rate coefficient in the
100-200 K range involves an  extrapolation.  Uncertain measured
activation energies for reactions 190 and 191 also lead to poorly
determined rate coefficients by 100-200 K.  The ion-molecule reaction
(740) may not even be exothermic, while the  rate coefficients for
the radiative association processes are order-of-magnitude estimates
at best.

\begin{table}
\caption{Additional important reactions for a perturbation amplitude of
1.0, for Model 2 and Composition A. }
\label{perturbation2}
         \begin{tabular}{l|l|rrr}
Number & Reaction$^{a}$ & $\alpha $&   $\beta $&   $\gamma $ \\
\hline
\hline
168 & C    +   SO$_2$ $\rightarrow $ CO +     SO & & & \\
173 & CH$_2$ +    O $\rightarrow $ CO  +    H  +     H & 1.20e-10  &
0.00  &    0.00 \\
308 & He$^{+}$ +  H$_2$O $\rightarrow $ OH  +  He  +  H$^{+}$ &
1.32e-09  & -0.50 & 0.00 \\
309 & He$^{+}$ + H$_2$O $\rightarrow $ OH$^{+}$ + He + H & 1.32e-09
&  -0.50  &    0.00\\
333 & He$^{+}$  +   OCS $\rightarrow $ CS + O$^{+}$ + He & 8.40e-10 &
-0.50 & 0.00\\
360 & C$^{+}$ + H$_2$O $\rightarrow $ HOC$^{+}$ +   H & 1.80e-09 &
-0.50   &   0.00\\
444 & CH$_{3}^{+}$ + S $\rightarrow $ HCS$^{+}$ + H$_2$ & & & \\
447 & CH$_{3}^{+}$ + SO $\rightarrow $ HOCS$^{+}$ +  H$_2$ & 4.20e-09
& -0.50 & 0.00 \\
459 & CH$_4$ + S$^{+}$ $\rightarrow $ H$_3$CS$^{+}$ +  H & 1.40e-10 &
0.00 & 0.00\\
509 & O + HCS$^{+}$ $\rightarrow $ S + HCO$^{+}$ & 5.00e-10 &
0.00 &     0.00\\
510 & O  +  HCS$^{+}$ $\rightarrow $ OCS$^{+}$ +   H & 5.00e-10 &
0.00 &     0.00\\
552 & H$_3$O$^{+}$ + H$_2$CO $\rightarrow $ H$_3$CO$^{+}$ +  H$_2$O &
2.60e-09 & -0.50 & 0.00\\
553 & H$_3$O$^{+}$ + CH$_4$O $\rightarrow $ CH$_5$O$^{+}$ +  H$_2$O & & & \\
646 & He$^{+}$ + H$_2$O $\rightarrow $ H$_2$O$^{+}$ + He & 1.32e-09 &
-0.50 & 0.00\\
663 & C$^{+}$ + S $\rightarrow $ S$^{+}$ + C & & & \\
\hline
\end{tabular}
\small{
\begin{list}{}{}
\item[$^{a}$]  Only rate coefficients that differ by more than 10
percent from the UMIST value are included.
\end{list}
}
\end{table}

\begin{table}
\caption{Additional important reactions at 200~K, for Model 2 and 
Composition A. }
\label{rates200K}
         \begin{tabular}{l|l|rrr}
Number & Reaction$^{a}$ & $\alpha $&   $\beta $&   $\gamma $ \\
\hline
\hline

107 & CO + CRPHOT $\rightarrow $ O + C & 1.30e-16 & 0.00 & 0.00 \\

125 & SO + CRPHOT $\rightarrow $ SO$^+$ + e$^-$ & 1.30e-14  &    0.00
&   0.00  \\

153 & H$_2$    + O      $\rightarrow $ OH  +  H & & &   \\

154 & H$_2$    +  OH    $\rightarrow $          H$_2$O +    H   &
8.40e-13&      0.00&   1040  \\

166 & C     +  SO    $\rightarrow $          S   +    CO        &
3.50e-11&      0.00&      0  \\

173 & CH$_2$   +  O     $\rightarrow $          CO   +   H   +    H
&  1.20e-10&      0.00&      0  \\

186 & O     +  OH    $\rightarrow $          O$_2$   +   H    &  &  &     \\

190 & O     +  H$_2$S   $\rightarrow $          HS   +   OH
&    9.22e-12&      0.00&   1800  \\

204 & OH    +  SO    $\rightarrow $          SO$_2$   +  H        & &
&        \\

309 & He$^+$ +    H$_2$O $\rightarrow $ OH$^+$ + He + H  & 1.32e-09
&   -0.50   &   0.00    \\

360 & C$^+$ + H$_2$O $\rightarrow $ HOC$^+$ + H   & 1.80e-09  &
-0.50   &   0.00    \\

436 & CH$_{3}^{+}$ +   O $\rightarrow $ HCO$^+$ + H$_2$ & 2.05e-10  &
0.00  &    0.00  \\

437 & CH$_{3}^{+}$ +   O $\rightarrow $ HOC$^+$ + H$_2$      & & & \\

646 & He$^+$ + H$_2$O $\rightarrow $ H$_2$O$^+$ +   He  & 1.32e-09 &
-0.50   &   0.00  \\

663 & C$^+$ + S $\rightarrow $ S$^+$ + C             &  &  &  \\

\hline
\end{tabular}
\small{
\begin{list}{}{}
\item[$^{a}$]  Only rate coefficients that differ by more than 10
percent from the UMIST value are included.
\end{list}
}
\end{table}

\section{Results}

\subsection{Abundances with respect to H$_2$}

\begin{figure}[h]
\begin{center}
\includegraphics[angle=90,width=\linewidth]{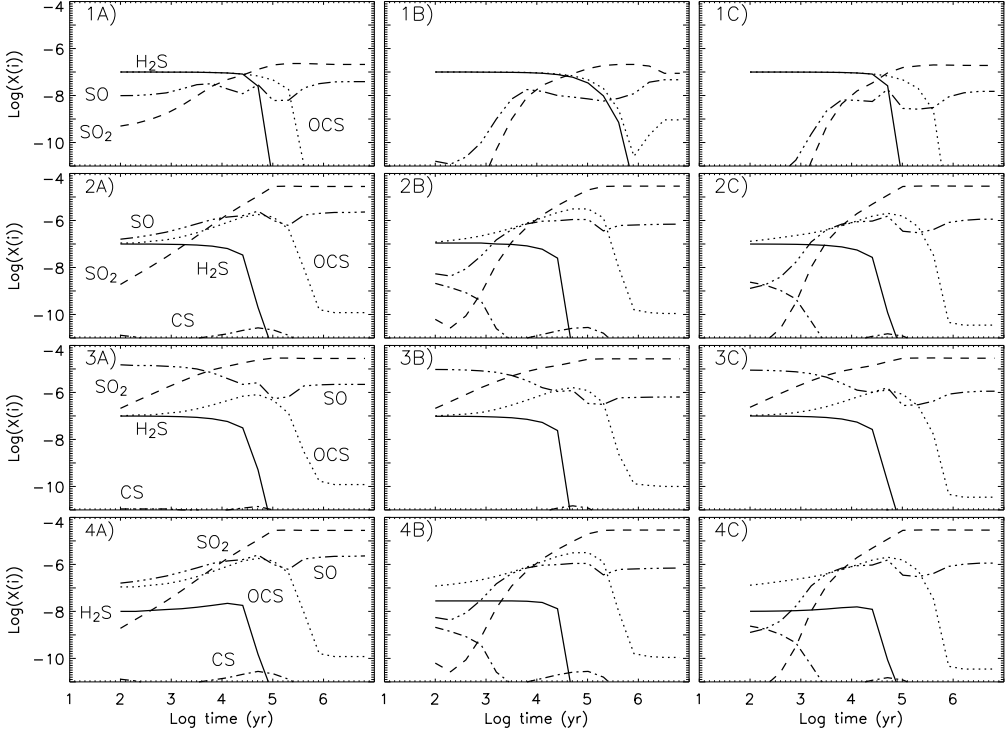}
\caption{Evolution of the SO$_{2}$, SO, H$_{2}$S, OCS and CS
abundances with respect to H$_2$ as a function of time, for the three
pre-evaporation gas phase
compositions A (left panels),  B (central panels), and C (right 
panels) and the four grain mantle compositions (1 to 4, from the top 
to the bottom). The
gas temperature is 100 K and the
H$_2$ density is 10$^7$ cm$^{-3}$.}
\label{3Comp}
\end{center}
\end{figure}

\begin{figure}[h]
\begin{center}
\includegraphics[angle=90,width=\linewidth]{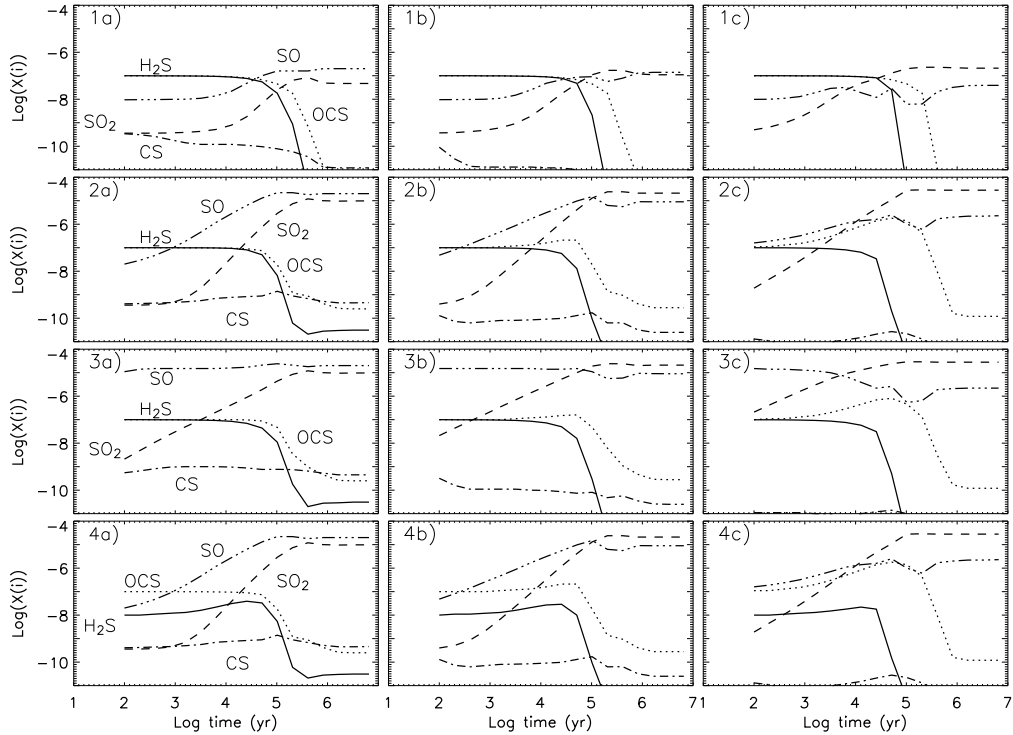}
\caption{Evolution of the SO$_{2}$, SO, H$_{2}$S, OCS and CS
abundances with respect to H$_2$ as a function of time, using 
pre-evaporated gas phase
composition A and the four grain mantle mixtures (1 to 4, from the 
top to the bottom).
The gas temperature is 100 K and the density is
$10^{5}$ cm$^{-3}$ (a, left panels), $10^{6}$ cm$^{-3}$ (b, central panels) and
$10^{7}$ cm$^{-3}$ (c, right panels).}
\label{plot100K}
\end{center}
\end{figure}

\begin{figure}[h]
\begin{center}
\includegraphics[angle=90,width=\linewidth]{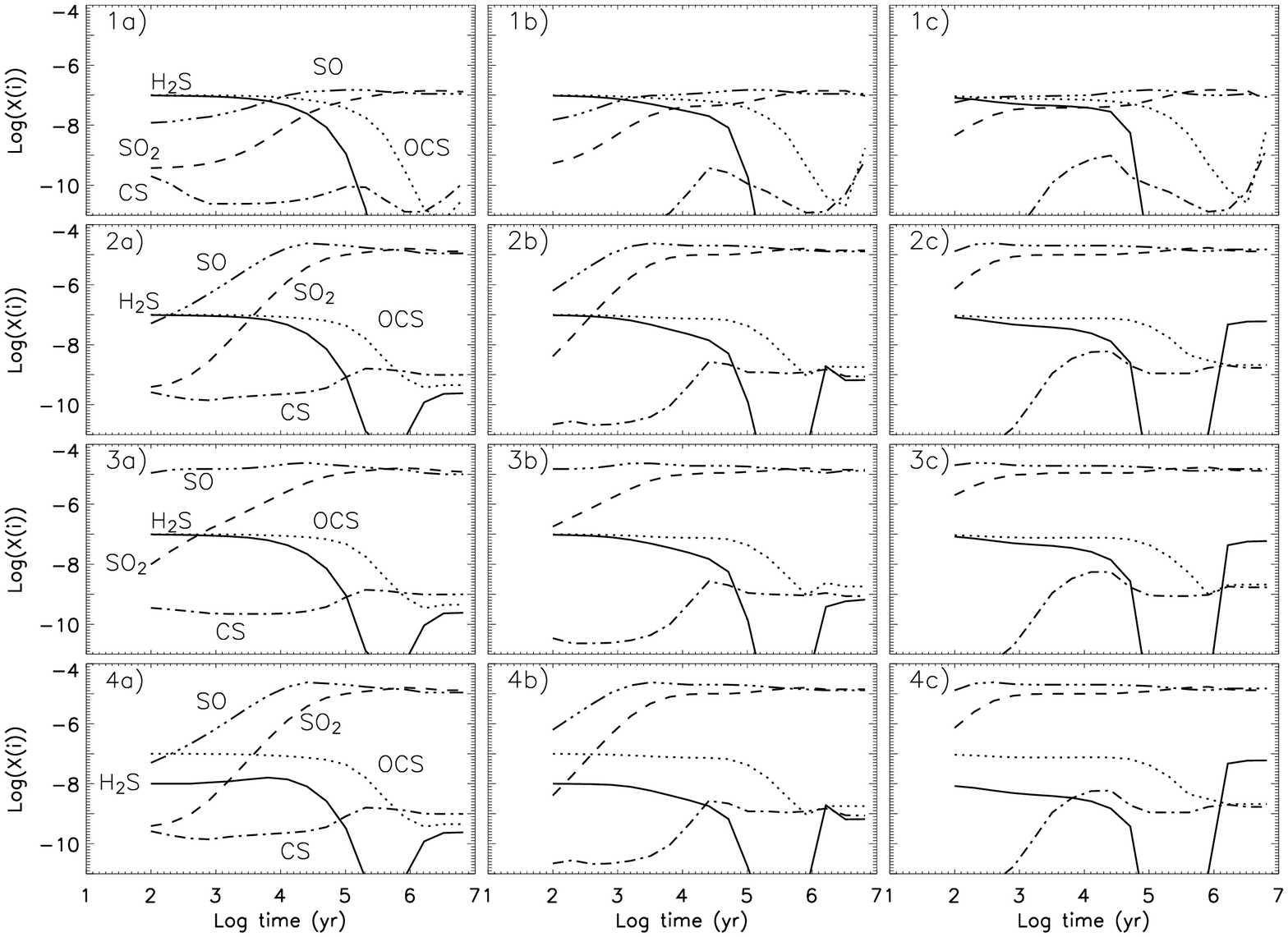}
\caption{Same as Fig.~\ref{plot100K} but with a temperature of 300 K.}
\label{plot300K}
\end{center}
\end{figure}

We have run the four models listed in Table \ref{iniabundS}, each with
      the three different gas-phase compositions (A, B, and C) prior to
evaporation, for gases with temperatures of 100 K and 300 K, and densities of
$10^{5}$~cm$^{-3}$, $10^{6}$~cm$^{-3}$ and $10^{7}$~cm$^{-3}$.  In 
this section, we give some sense of the major
features of the chemistry of the sulphur-bearing species.

Fig.~\ref{3Comp} shows the evolution of the
abundances of the SO$_{2}$, SO, H$_{2}$S, OCS and CS molecules, with
respect to H$_2$, for the
four models and three gas-phase compositions (with T=100~K and n(H$_2$)=$10^{7}$~cm$^{-3}$).  
There are clearly differences among 
the models, at times earlier than
$10^4$~yr, due to the different initial
compositions whereas after $10^4$~yr the three gas phase compositions give
similar results.  For example,
for Models 1, 2 and 4, the compositions B and C give lower SO and
SO$_2$ abundances than the composition A at times earlier than $10^4$~yr,
whereas the three gas phase compositions give similar results for Model 3.
For sake of simplicity, in the following, we will discuss composition A, but
the results do not change qualitatively assuming the B or C compositions.

Figs.~\ref{plot100K} and \ref{plot300K} show the evolution, at
100 K and 300 K respectively, of the
abundances of the
main S-bearing species for the four grain mantle compositions of 
Table~\ref{iniabundS} at the three different densities.
In all the models depicted, SO and SO$_2$ are the most abundant species at late
times. The final large amounts of SO$_2$ are more noticeable in 
Models 2-4, where large amounts of
gaseous sulphur are available.
At a temperature of 100~K and a density between $10^{6}$ and
$10^{7}$~cm$^{-3}$,
the SO$_2$ molecule becomes more abundant than SO after
$10^4$-$10^5$~yr, mainly
because of the neutral--neutral reaction 
O~+~SO~$\rightarrow$~SO$_2$~+~photon.  Note that this radiative
association reaction is critical because of the high abundance of 
atomic O in the pre-evaporative gas.
At 300 K, the SO$_2$ molecule is formed less efficiently via this 
mechanism since it possesses an inverse
dependence on temperature
(see Table~\ref{rates}, reaction 756) but it is still as abundant
as SO after
$2\times 10^5$~yr for a density of $10^{5}$~cm$^{-3}$, and after
$10^3$~yr for a
density of $10^{7}$~cm$^{-3}$ (see Fig.~\ref{plot300K}). At 300 K,
OH is quickly ($\leq 10^2$~yr) formed through the reaction
H$_2$~+~O~$\rightarrow$~OH~+~H so that SO$_2$ can be  formed by
the reaction OH~+~SO~$\rightarrow$~SO$_2$~+~H. Here, the presence of a
large atomic oxygen
abundance in the pre-evaporated gas-phase is crucial to produce the 
high abundance
of OH at early
times,
contrary to what was found in previous models \citep[e. g.
][]{1997ApJ...481..396C}.
In Model 3, where the initial sulphur is mostly in S$_2$, the SO
molecule is very quickly ($\leq 10^2$~yr) formed, as S$_2$ directly
leads to SO through the
reaction S$_2$~+~O~$\rightarrow$~SO~+~S.

Now, let us look at the chemistry of hydrogen sulphide, OCS, and CS. The
initial H$_2$S abundance (see Table~\ref{iniabundS}),  dips sharply after $10^4$~yr in all
models but increases after
$10^6$~yr for models 2 to 4 at 300~K. The decrease of the H$_2$S abundance
at $10^4$~yr
is produced by the attack on H$_2$S by
H$_3$O$^{+}$, more abundant than H$_3^+$ in regions, such as hot cores,
where water has a large abundance.  The reaction between H$_3$O$^{+}$ 
and H$_2$S yields
protonated hydrogen sulphide (H$_3$S$^+$), which dissociatively
recombines with electrons to form HS at least part of the time:
H$_3$S$^{+}$~+~e$^{-}$~$\rightarrow$~HS~+~H~+~H.   The HS product is 
itself depleted by the reaction
HS~+~O~$\rightarrow$~SO~+~H.
At 300 K, H$_2$S is efficiently formed at later times ($\ge 10^5$~yr) by a
series of  reactions that starts with the destruction by cosmic
ray-induced photons  of SO and SO$^+$ to produce atomic sulphur.
Atomic sulphur is then hydrogenated into HS first, and then H$_2$S
(with intermediate steps in which the unusual species HS$_{2}^{+}$, 
H$_3$S$_{2}^{+}$,
HS$_{2}$ and H$_2$S$_{2}^{+}$ are formed).
Note that at lower temperatures S is  oxygenated rather than hydrogenated,
whereas at 300 K atomic oxygen goes into water and, therefore, S can 
be hydrogenated eventually.

As in the case of H$_{2}$S, the initial (adopted) abundance of OCS
is that derived from observations of this species in the solid state
(see Section 2.1 and Table 2). Once in the gas phase, OCS
is destroyed
later than H$_2$S. Under some conditions, e.g. T~=~100~K and $\rm n({H_2})$ =
10$^6$--10$^7$ cm$^{-3}$,
this molecule maintains a sizable if reduced abundance.  Actually, at 
T~=~100~K and
  $\rm n({H_2})$~=~10$^6$--10$^7$ cm$^{-3}$, there is even a temporary 
increase in the OCS abundance from $\sim$ 10$^{-7}$ to $\sim$ 
10$^{-6}$  (see
Figure~\ref{plot100K}). The OCS molecule is mostly formed by the 
radiative association reaction CO + S $\rightarrow$ OCS + PHOTON and 
this reaction is aided by the high density.

Unlike the other species, the CS molecule does not start on grain 
mantles in our calculations, but is part of the pre-evaporative gas. 
For composition A, its initial abundance is rather low.
At 100 K, the initial CS is destroyed increasingly efficiently as
  density increases  and its abundance never goes over $10^{-9}$. At 
300 K,  CS is efficiently produced at high density
after $10^3$~yr and it
can be as abundant as $10^{-8}$.    For compositions B and C, there 
is significantly more CS present initially, especially for 
composition B, which contains 50 times more CS than in composition A. 
The initial CS is more slowly destroyed
using compositions B and C
than using composition A under the conditions depicted in 
Fig.~\ref{3Comp} for Models 2 and 4.

\subsection{Abundance ratios}

\begin{figure}[h]
\begin{center}
\includegraphics[width=\linewidth]{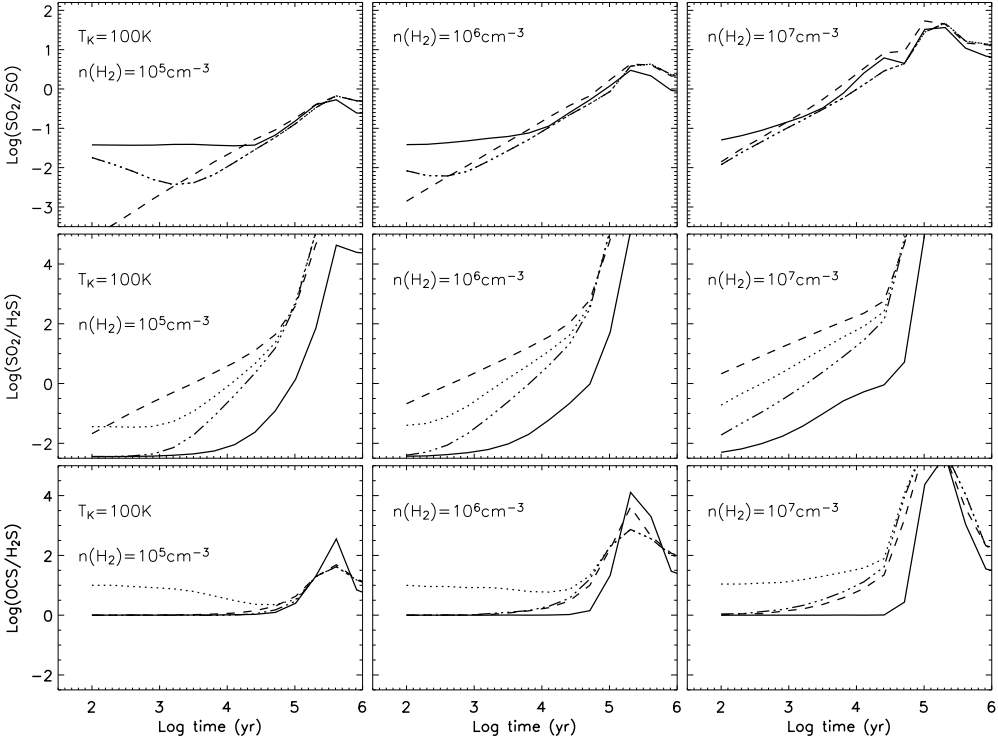}
\caption{Evolution of the abundance ratios SO$_{2}$/SO (upper panels),
SO$_{2}$/H$_{2}$S (middle panels) and OCS/H$_{2}$S (lower panels)
  as  functions of time for composition A,
a gas temperature of 100 K, and  densities of
$10^{5}$ cm$^{-3}$ (left panels), $10^{6}$ cm$^{-3}$ (central panels) and
$10^{7}$ cm$^{-3}$ (right panels). The solid, dashed-dotted, dashed, 
and dotted lines represent
results from Models 1,2,3, and 4, respectively.
In the upper panels, the results of Model 2 and and 4 are the same
and represented by dashed-dotted lines.}
\label{ratio100K}

\end{center}
\end{figure}
\begin{figure}[h]

\begin{center}
\includegraphics[width=\linewidth]{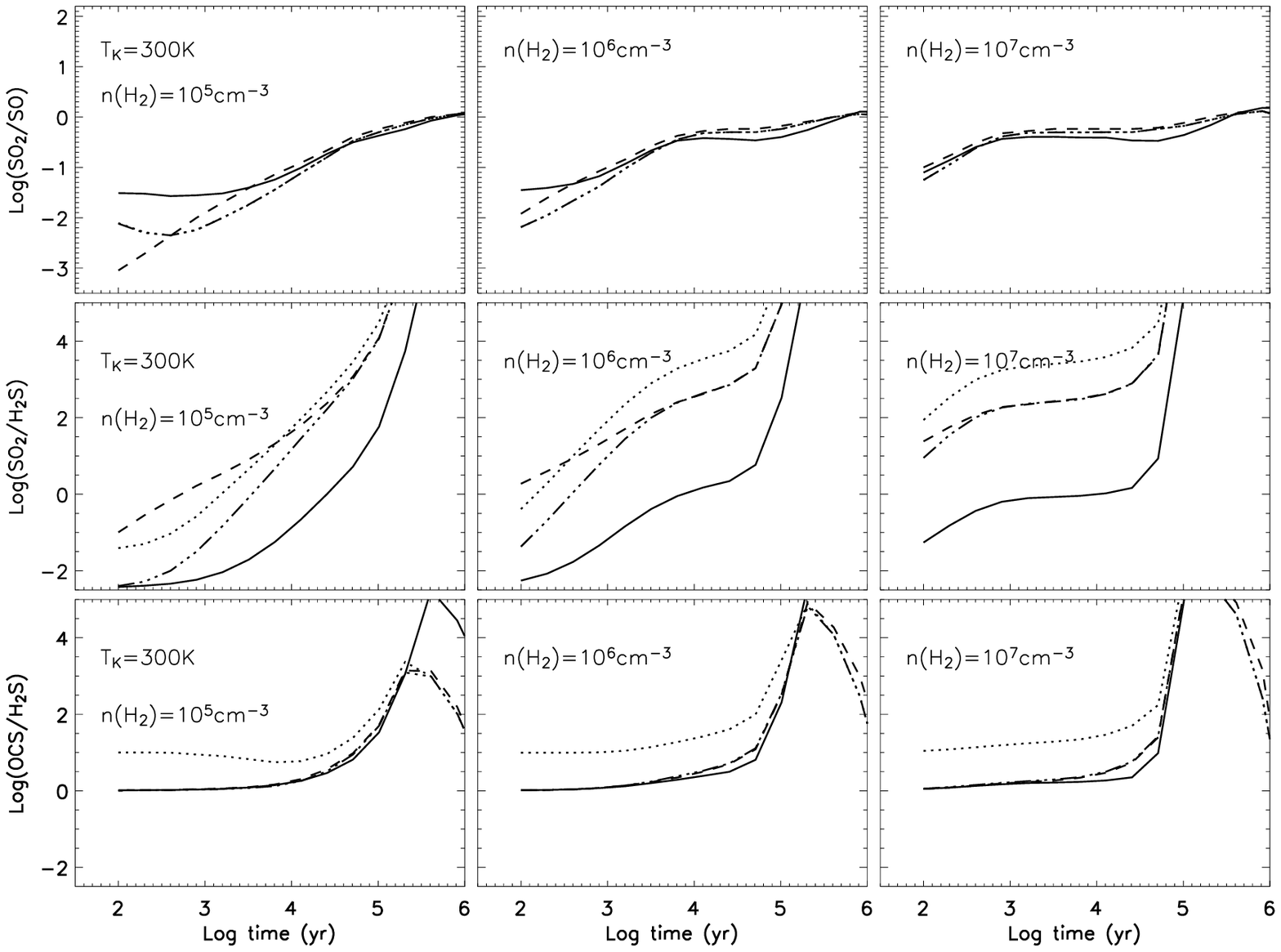}
\caption{Same as Fig.~\ref{ratio100K} but with a temperature of 300 K.}
\label{ratio300K}
\end{center}
\end{figure}
\begin{table}
\caption{Maximum sensitivity of the abundance ratios to assorted parameters. }
\label{max_dev}
\begin{tabular}{llrl}
\hline
Ratio & Changing parameter & MRV & Set of parameter \\
\hline
\hline
SO$_2$/SO & Time & 16 & 100~K, $10^7$~cm$^{-3}$, Mod.~3\\
& Temperature & 6 & $10^4$~yr, $10^5$~cm$^{-3}$, Mod.~4\\
& Density & 125 & $10^4$~yr, 100~K, Mod.~3\\
& Model & 20 & $10^3$~yr, 100~K, $10^5$~cm$^{-3}$\\
\hline
SO$_2$/H$_2$S & Time & 330 & 300~K, $10^5$~cm$^{-3}$, Mod.~2\\
& Temperature & 630 & $10^3$~yr, $10^7$~cm$^{-3}$, Mod.~4\\
& Density & 5000 & $10^3$~yr, 300~K, Mod.~4\\
& Model & 3150 & $10^4$~yr, 300~K, $10^7$~cm$^{-3}$\\
\hline
OCS/H$_2$S & Time & 6 & 100~K, $10^7$~cm$^{-3}$, Mod.~2\\
& Temperature & 3 & $10^4$~yr, $10^7$~cm$^{-3}$, Mod.~2\\
& Density & 10 & $10^4$~yr, 100~K, Mod.~2\\
& Model & 30 & $10^4$~yr, 100~K, $10^7$~cm$^{-3}$\\
\hline
\end{tabular}
\end{table}
Abundance ratios between sulphur-bearing species are exceedingly important
  because the use of fractional
abundances ($X(i) = N(i)/N(H_{2}$))
as chemical clocks introduces an additional parameter in the analysis -- the
H$_{2}$ column density (N(H$_{2}$)) in the emitting region -- which is rarely
well constrained observationally.  Moreover, the abundance ratios are
less sensitive to the
initial amount of sulphur species compared with absolute abundances.
For example, an initial
abundance of S or S$_2$ two times less than those used in Model 2 or
3 does not change significantly the abundance ratios. This
is why observed abundance ratios are typically used to put contraints
on chemical models, when the two species trace the same region. 

Figs.~\ref{ratio100K} and \ref{ratio300K} show the evolution of the
abundance ratios SO$_{2}$/SO, SO$_{2}$/H$_{2}$S and OCS/H$_{2}$S,
at 100 K and 300 K for three densities using composition A and all four models.
The overall sense of these figures is that
      the computed ratios are more sensitive to the gas temperature, 
the density,
and  the (poorly known) form of sulphur in grain mantles than to the time
(particularly for $t$ $\leq 3 \times 10^4$ yr) or to the gas phase
chemical history of the cloud. As a first step in quantifying the
sensitivity of the studied abundance ratios to the different  parameters,  we report in Table~\ref{max_dev} the 
maximum relative variation (MRV) for the three abundance ratios. These are obtained by changing the parameters time, temperature, density, model within the following considered ranges: time, 10$^3$-10$^4$~yr; temperature, 100-300 K; density, 10$^5$-10$^7$~cm$^{-3}$; grain mantle composition, Models 1-4. The MRV  is determined for each parameter by varying this parameter within the stated range while holding the others fixed at the set of values shown in Table~\ref{max_dev}. These sets lead to the strongest changes; hence the term "maximum relative variation". For example, the maximum variation of the SO$_2$/SO ratio produced by an increase of the time from $10^3$ to $10^4$~yr is 16 and it is obtained for a temperature of 100~K, a density of $10^7$~cm$^{-3}$ and Model 3.

The SO$_{2}$/H$_{2}$S ratio shows the largest variations with time,
but, unfortunately, also with differing physical conditions
and mantle mixtures. Indeed the ratio is more sensitive to the latter parameters
than to the time, for $t \leq 3 \times 10^4$ yr.
The OCS/H$_{2}$S ratio is
less sensitive to the different mantle mixtures and
physical conditions, except at late times ($\geq 10^{5}$ yr).
Unfortunately, however, this ratio is not sensitive to the time
at $t \leq 10^{5}$ yr either.
The SO$_{2}$/SO ratio shows significant variations with respect
to the different mantle compositions, and different densities
and temperatures at relatively early times ($\leq 10^{4}$ yr).
These variations mask completely the dependence on the time,
which is, anyway, moderate at 100 K with the exception of the high density
case, and small at 300 K.

 To gain an  understanding of the sensitivities
of these abundance ratios in addition to that obtained from the MRV analysis, we follow the variations of
SO$_{2}$/SO, SO$_{2}$/H$_{2}$S and OCS/H$_{2}$S  compared with
reference ratios computed from Model 2 at 10$^{3}$~yr for a
temperature of 100~K and a density of $10^{5}$~cm$^{-3}$ (left panels
of Fig.~\ref{ratio100K}). The reference SO$_{2}$/SO ratio is
$\sim 5\times 10^{-3}$ and increases by only a factor of 2 by
$10^{4}$~yr.
However, a similar increase can also be due to an underestimate of the
density by a factor of 10  (note that if $n_{\rm H_{2}}$ = 10$^7$ cm$^{-3}$,
the SO$_{2}$/SO ratio increases by a factor of 20). Also, the reference SO$_{2}$/SO ratio
at 10$^3$ yr varies by a factor of $\sim$20 depending on the adopted
mantle mixture. The
reference SO$_{2}$/H$_{2}$S ratio is 0.01 and increases by one order
of magnitude at $10^{4}$~yr. An increase in density (to
$10^{7}$~cm$^{-3}$) or in temperature (to 300~K) gives respectively a
ratio of 0.3 or 0.1. The largest variation is seen if both the
temperature and the density increase, so that the SO$_{2}$/H$_{2}$S
ratio jumps to 160. A different mantle mixture leads to ratios
between $3\times 10^{-3}$ and 0.3 (at the reference time).
The reference OCS/H$_{2}$S ratio is 1 and shows little to no
sensitivity to time until $10^4$~yr,  but also evinces
little
change with temperature, density and mantle composition,
with the exception of Model 4 which gives a ratio enhanced by a factor
of $\sim$10 because the abundance of H$_{2}$S is 10 times less than in
Models 1, 2, and 3.

\section{Discussion}

\subsection{Comparison with previous models}

The most recent papers focussing on the detailed modelling of sulphur 
chemistry in
hot cores are those of
\citet{1997ApJ...481..396C} and \citet{1998A&A...338..713H}. The adopted
chemistry is roughly the
same in the two models, but they differ in the value of the activation energy
barrier for the destruction
of H$_2$S by atomic hydrogen (H$_2$S + H $\rightarrow$ HS + H$_2$).
In particular, \citeauthor{1998A&A...338..713H} used a value of 350~K 
while \citeauthor{1997ApJ...481..396C}
used a value of 850~K.
Although there is some experimental support for a variety of choices, 
here we are using an energy
barrier of 1350 K, based on the recent experiment by
\citet{1999Peng} in the 300-600~K range.
 The difference in the activation energy barrier
 for the H$_2$S + H reaction causes significant variations in the models.
While we obtain results similar to the Charnley model for 100~K,
H$_2$S destruction occurs at later times in our model for T~=~300~K.
Moreover, the high value of the energy barrier has significant 
consequences on the SO and SO$_2$ abundances of our model with 
respect to the above two models. For example
at $10^4$~yr, our model predicts SO and SO$_2$
abundances 6 and 30 times lower than \citeauthor{1997ApJ...481..396C}'s model, using
the same parameters and initial conditions.
Overall, and in addition to this, our results are different from 
these two models primarily because
we are assuming different initial gas phase and/or mantle compositions.
As already noted (in section 3.1), the large atomic oxygen abundance 
is, for example,
an important difference in the pre-evaporated gas phase composition.
Another difference is that both \citeauthor{1997ApJ...481..396C} and 
\citeauthor{1998A&A...338..713H}
assume that the bulk of sulphur in the ice mantle is in the H$_2$S form.
Our model 1 adopts a similar mantle mixture, with solid-phase OCS as 
abundant as H$_2$S,
whereas models 2, 3 and 4 assume that the bulk of the sulphur lies in 
the S or S$_2$ forms.
These two classes of models give very different results, as widely 
discussed in section 3.

Actually, a novelty of this work is the consideration that the 
sulphur can be in the
atomic or molecular form when evaporated from the grain mantles.
However, we should emphasize that atomic S is very quickly locked 
into SO and SO$_2$
  molecules, so that we do not expect large atomic S quantities in the gas phase
for a long time.
For example, for a gas at 300 K and $10^7$~cm$^{-3}$, the S abundance decreases
from $3\times 10^{-5}$ to $1.5\times 10^{-9}$ in $10^3$~yr by reacting with O$_2$ to form SO.
This explains why \citet{2003A&A...412..133V} did not detect atomic 
sulphur\footnote{\citet{2003A&A...412..133V} argued that the non 
detection
of the 25 $\mu$m line corresponds to an upper limit to the atomic sulphur of
$\leq 5\times 10^{-8}$, computed by considering line absorption.
Indeed, this is an upper limit on the S abundance in the absorbing 
gas, i.e. the
foreground cold gas, rather than the hot gas. The upper limit on the 
emitting hot core
gas is much higher, as explained in the text.} in
the hot core region of the massive protostars that they studied, where the gas
temperature and density are similar to those quoted above.
In a slightly colder gas ($\sim 100$ K), atomic S would survive longer, but it
would be very difficult to detect.
In fact the intensity of the SI fine structure line at 25.249 $\mu$m is
$1\times 10^{-12}$ erg s$^{-1}$ cm$^{-2}$, for a gas at 100 K and 
$10^7$ cm$^{-3}$, a source
size of $2"$, an H$_2$ column density of $10^{23}$~cm$^{-2}$ , and all
the sulphur in  atomic form (as our models 2 and 4 adopt). The signal would be
attenuated by the foreground dust of the cold envelope, whose H$_2$ column density is also
around $10^{23}$ cm$^{-2}$, by about a factor of ten \citep[using the
extinction curves in ][]{2003ARA&A..41..241D} putting the signal at
the limit of the ISO-SWS detection (a typical rms is few times $10^{-13}$ erg s$^{-1}$ cm$^{-2}$). 
Future investigations in the ISO database will be done in order to check the 
possible presence of atomique sulphur in hot cores.

\citet{1998A&A...338..713H} found the cosmic ray rate to be an
important parameter changing the
timescales of the destruction of H$_2$S and the formation of SO and
SO$_2$. The standard value usually
assumed is $1.3\times 10^{-17}$ s$^{-1}$ but there are some
indications that it can vary depending on the
region \citep[see ][]{2000A&A...358L..79V}.
To check the dependence on ionization rate whether or not it can 
pertain to objects as dense as hot
cores, we have run the models for three different cosmic ray
ionization rates:
$1.3\times 10^{-17}$, $1.3\times 10^{-16}$ and $1.3\times 10^{-15}$~s$^{-1}$.
The results for the S-bearing abundance ratios do not strongly depend 
on this parameter at a
density of $10^{5}$~cm$^{-3}$.
At higher densities, however,  an increase in
the cosmic ray ionization rate speeds up the photodestruction of OCS, H$_2$S and SO$_2$ via cosmic ray-induced photons.
 For a cosmic ray ionization rate of $1.3\times 10^{-15}$~s$^{-1}$, a 
temperature of 100~K
and a density of $10^{7}$~cm$^{-3}$,
the abundances of H$_2$S and OCS start to decrease before $10^3$~yr
and the maximum abundance of SO$_2$ is
$4.5 \times 10^{-8}$ instead of $2 \times 10^{-7}$ with Model 1 and 
composition A. Hence, as \citet{1998A&A...338..713H}, we found that the adopted value of the cosmic ray rate may be an important parameter at high density whereas it has weak consequences on the results at lower densities.

 Recently, \citet*{2003nomura} reported a study of the chemical
composition across the envelope of the massive protostar G24.3+0.15. 
In this study, they derived the density and temperature profile and computed
the chemical composition of the gas as a function of the radius and time. 
Evidently, the \citet*{2003nomura} model
is more complete than ours in dealing with the physical structure of the
protostar, for this is the focus of their model. On the other hand, given its
complexity, the model does not explore the robustness of the achieved results
as a function of the necessary assumptions of the model itself, which is, on
the contrary, the focus of our study. In 
the same way, \citet{2002A&A...389..446D} studied the chemical 
composition across the envelope of the massive protostar AFGL 2591. Finally, a further degree of complexity has been added to the problem by \citet{2003ApJ...585..355R}, who considered the evolution of a protostar,
including the evolution of the thermal and physical structure of the envelope
plus the chemical evolution. As in the previous case, the advantage of having a better description of the protostar comes along with the disadvantage of a lack of exploration of the
robustness of the results as a function of the chemical network utilized. In these three models, the authors assumed that sulphur was only evaporated from the grains mantles in the OCS and H$_2$S forms for \citet*{2003nomura} and only in the H$_2$S form for \citet{2002A&A...389..446D} and \citet{2003ApJ...585..355R}. 
We have discussed above the problem with the assumption that sulphur is totally frozen 
onto OCS and H$_2$S, and showed that the resulting S-bearing 
abundances depend dramatically on this assumption. The conclusion is the same if considering a model where sulphur is only in H$_2$S (as discussed at the end of Sect.~4.2). It comes as no surprise, therefore, that our results are different from those of these more detailed approaches.

\subsection{Comparison with observations}

In the following, we consider the case of two well studied hot cores: 
the massive Orion KL and the
low mass IRAS16293-2422 hot cores. Orion KL is a complex region
where several energetic phenomena are present. 
High resolution, interferometric observations have
shown that different molecules originate in different components, especially for the OCS, SO and SO$_2$ molecules 
\citep[e.g. ][]{1996ApJ...469..216W}.
It is evidently a very crude 
approximation to attribute all the
S-molecule emission to the hot core, but lacking a better
understanding of the region
we will try to compare the observed abundances with our model predictions.
First, in order to minimise the number of  parameters, we will 
compare available observations
with computed abundance ratios. The SO, SO$_2$, H$_2$S and OCS 
column densities have been derived by \citet{2001ApJS..132..281S} and \citet{1995ApJS...97..455S} to be
$2.3 \times 10^{17}$, $6 \times 10^{16}$, $1.2 \times 
10^{16}$~cm$^{-2}$ and $9 \times 10^{15}$~cm$^{-2}$, respectively.
All these column densities are beam-averaged, with a beam of $\sim 10''$.
For simplicity, we will assume that the emission originates in the 
hot core region with
T~=~200~K and n(H$_2$)~=~$10^{7}$~cm$^{-3}$ \citep{1992ApJ...393..225W}.
The comparison of the abundance ratios derived from the above column densities
with our model predictions shows that none of the four models of 
Sect.~2.1 reproduces the
observed data.
However, the ratios can be reproduced by a model similar to Model 2,
where the initial abundance of atomic sulphur injected into the
gas phase is $3\times 10^{-6}$, i.e. ten times lower than the 
abundance used in Model 2.
We will call this model Model 2'.
In that case, we obtain a good agreement between the model and the observations and
derive an age of $4\times 10^3$~yr.
Assuming the emission region extends about $10"$ and an H$_2$ column
density of $8 \times 10^{23}$~cm$^{-2}$ \citep{1995ApJS...97..455S},
 the observed abundances of SO, SO$_2$, H$_2$S and OCS are reproduced to within a factor of 10. On the contrary, the CS abundance is underestimated by three orders of magnitude by our model, probably because
CS emission does not originate in the hot core. Note that, given the complexity of the region, a better analysis would require an understanding of the H$_2$ column density
associated with the S-bearing molecules.

\begin{table}
\caption{Observed and modelled abundances of SO, SO$_2$, H$_2$S, OCS
and CS towards the IRAS16293-2422 hot core.
\label{irasab}}
         \begin{tabular}{l|ccc}
\hline
Species & observed & Model 2 & Model 2' \\
\hline
\hline
SO	& $1.7\times 10^{-6}$ & $5\times 10^{-7}$ & $4\times 10^{-7}$ \\
SO$_2$	& $5.4\times 10^{-7}$ & $10^{-7}$ & $10^{-7}$ \\
H$_2$S	& $5.3\times 10^{-7}$ & $10^{-7}$ & $10^{-7}$\\
OCS 	& $1\times 10^{-6}$ & $2\times 10^{-7}$ & $10^{-7}$ \\
CS	& - & $1\times 10^{-11}$ & $2\times 10^{-12}$ \\
\hline
\end{tabular}
\end{table}

\begin{figure}[h]
\begin{center}
\includegraphics[width=9cm]{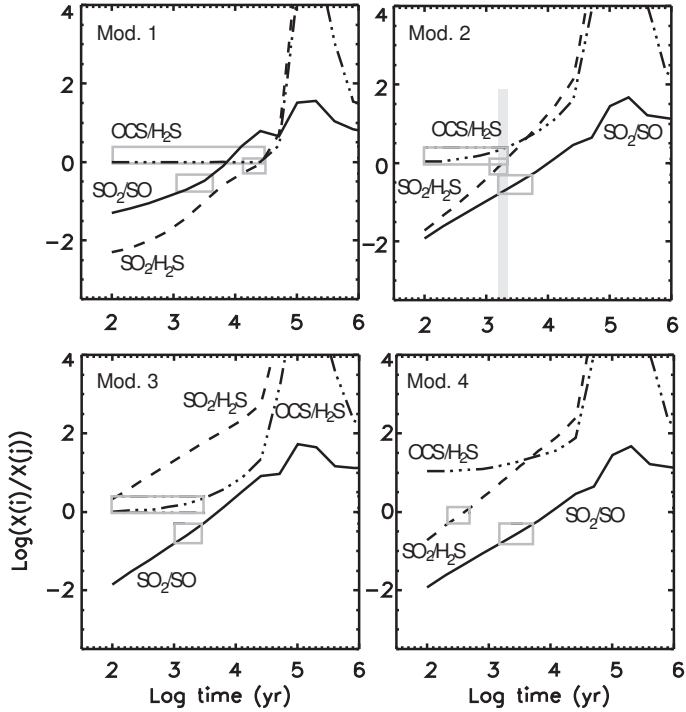}
\caption{Comparison between the observed ratios SO$_2$/SO,
SO$_2$/H$_2$S and OCS/H$_2$S towards
IRAS16293-2422 and the theoretical predictions of four models (Mod.
1, 2, 3 and 4; all with composition A) for a gas temperature of 100~K and a
density of $10^7$~cm$^{-3}$.  The grey empty boxes, if shown, 
represent those instances in which particular calculated ratios are in
agreement for a range of times with the observed ones
taking into account the uncertainties in the observations. The grey 
filled box for Model 2 shows the time interval where all calculated 
ratios are
in agreement with the observed ones.  }
\label{4ratio}
\end{center}
\end{figure}

In the case of the low-mass protostar IRAS16293-2422,
  multifrequency observations of H$_2$S, OCS, SO and SO$_2$
have been reported and analyzed by 
sophisticated models, which take into account
the density and temperature structure of the source, as well as
the abundance profile of each studied molecule \citep{2002A&A...390.1001S,2004A&A...413..609W}.
Both \citet{2002A&A...390.1001S} and \citet{2004A&A...413..609W} 
found the following abundance ratios in the inner warm region where the dust mantles 
evaporate \citep{2000A&A...355.1129C}:
SO$_{2}$/SO=0.3, SO$_{2}$/H$_{2}$S=1 and OCS/H$_{2}$S=1.9.

 The modelling of the density and temperature profiles of
IRAS16293-2422, by \citet{2000A&A...355.1129C}, suggests a density of
$10^7$~cm$^{-3}$ and a temperature
of 100~K in the hot core region. Fig.~\ref{4ratio} shows the 
comparison between the observed
  and predicted ratios (assuming composition A) for the four models: 
only Model 2 reproduces the
observations, suggesting an age of $\sim 2\times 10^3$~yr for the protostar.
To check  the robustness of this result, we also tried a variety of 
intermediate mantle mixtures, between Model 2 and Model 3, with the 
following characteristics:
(1) one-half of the initial mantle sulphur in atomic form and one-half in
the form of S$_{2}$, (2) one-quarter in S and three-quarters
in S$_2$, (3) three-quarters in S and
one-quarter in S$_2$. In all three intermediate
cases, the resulting curves are very similar to those found for Model 
3. The presence
of molecular sulphur clearly determines the
form of the curves and is inconsistent with these observations.
The observed ratios are, however, well reproduced by Model 2' (used 
for the comparison with Orion) giving an age, $\sim 3\times 10^3$~yr,
very similar to the age derived by Model 2.
Finally, both Model 2 and Model 2' predict fractional abundances at 
the optimum times in good
agreement (within a factor of 5-10) with those observed, as reported 
in Table~\ref{irasab}. Note that the CS abundance towards 
IRAS16293-2422 has not been derived 
\citep{2002A&A...390.1001S,2004A&A...413..609W} and that the 
abundances of the main S-bearing species predicted by Model 2 and 2' 
are very similar until $10^4$~yr.
This comparison suggests that the majority of the sulphur is released into
the gas phase in its atomic form or quickly (t $\leq 10^{3}$~yrs)
converted to it, and that the abundance of the H$_{2}$S molecule injected in
the gas phase from the grains mantles cannot be much less than $10^{-7}$.

The age determined for IRAS16293-2422 is relatively short compared 
with the previous estimates ($\sim 10^4$~yr) 
\citep{2000A&A...357L...9C,2002A&A...395..573M,2004A&A...413..609W}.
However, our newly determined age
represents only the time from the evaporation of the mantles;  the dynamical
time needed to reach the required luminosity to form the observed hot core is
an additional $\sim 3\times 10^4$~yr.

The discussion of the best chemical age for IRAS 16293-2422 is based 
on models with the
standard value of the cosmic ray ionization rate ($1.3 \times 10^{-17}$~s$^{-1}$). If
we use a rate 10 times higher as suggested by \citet{2004doty}, the observations towards
IRAS16293-2422 are no longer in agreement with
Model 2, but with Model 1, although the derived age is very similar. 
Indeed, the enhanced cosmic ray ionization rate speeds up the 
destruction of H$_2$S, increasing  the ratio SO$_2$/H$_2$S more
rapidly without affecting significantly the other ones. Consequently, 
the  SO$_2$/H$_2$S curves on Fig.~\ref{4ratio} are shifted to the 
right of the figures, worsening the agreement with Model 2 but 
improving it with Model 1. In that case, the derived age is $5 \times 10^{3}$~yrs and the predicted absolute abundances are between 15 and 20 less than the observed ones, however. Thus, a high cosmic ray ionization rate is no longer compatible with our hypothesis that mantle evaporation leads quickly to high gas-phase abundances of atomic
sulphur. The high cosmic ray ionization rate has also a consequence on the O$_2$ abundance pedicted to be $\sim 10^{-6}$ at $10^3$~yr with Model 1 whereas it is predicted to be two orders of magnitude lower with the standard value of the cosmic ray ionization rate and Model 2. However, at present one cannot choose between the two conditions (large abundances of S in the gas phase or large cosmic-ray ionization rate).  High sensitivity observations of atomic sulphur (see Sect. 4.1) are needed to put stringent constraints on chemical models.

Another way to confirm our hypothesis of a 
large initial abundance of atomic sulfur would be a careful study involving several other protostars at different stages
of evolution.
  The strongest prediction of our models 2 to 4 is the large
($\ga$ 10$^{-5}$)
amount of SO$_2$ at late times, compared with
models that start with little sulphur evaporated from the grains. There is some
evidence for SO$_2$ abundances in high-mass hot cores as large as $10^{-6}$
\citep{2003A&A...412..133V},
suggesting that an initial amount of sulphur at least higher than $10^{-6}$
is needed. But our prediction of  large SO$_2$ abundances in
more evolved hot cores with ages larger than $\sim$ 10$^{5}$~yr, does 
not pertain to the
  Orion and IRAS16293-2422 hot cores (as found in the
present work).

Quantitative comparisons with  observations of other hot
cores reported in the literature,
such as the observations by \citet{1998A&A...338..713H} and
\citet{2003A&A...399..567B},
are difficult to carry out since the abundances determined by these authors are averages along
the line of sight and beam-averaged. Hence, their analyses do not take into
account the physical
structure of the source.
Finally, \citet{2003A&A...412..133V} report the study of the 
S-bearing molecular
abundance in half a dozen massive protostars. In this case, an attempt to
disentangle the outer envelope
and inner core has been done, but the analysis is not accurate enough to derive
abundance ratios to compare with detailed model predictions; rather, 
the authors  just give
order-of-magnitude jumps of abundance in the inner hot core regions.
\citet{2003A&A...412..133V} compared the measured abundances with the model
predictions by \citet{2002A&A...389..446D}, and argued that OCS is 
the main sulphur bearing molecules on the mantles, because its predicted abundance is otherwise
too low.
We ran a model where sulphur is in solid OCS, to test this 
suggestion, and we found that
it leads to results very similar to those of Model 1 (where sulphur is evaporated from the grain mantles in the H$_2$S and OCS forms), because
OCS has almost the same chemical behavior as H$_2$S. Although OCS is
destroyed at later times, the differences in the SO and SO$_2$
formation are not significant and the same discussion about Model 1 is valid.

\section{Conclusions}


We have studied in detail the influence of the mantle form of
sulphur on the post-evaporative gas-phase abundances of S-bearing
molecules in hot star-forming regions,
with the goal of understanding whether those molecules can be used
to estimate ages.
We considered four different reasonable mantle mixtures,
from which gas-phase
     H$_2$S, OCS, S and S$_2$ emerge after a process of evaporation and,
    for the last two species,  possible rapid reaction, with different
relative
abundances, joining other species in the gas-phase prior to 
evaporation. We then followed the post-evaporative chemical 
evolution, with an emphasis on the abundance ratios of the main
sulphur-bearing species for realistic physical conditions present in
hot cores.
Our results show that none of the ratios
involving the four most abundant
S-bearing molecules, namely H$_2$S, OCS, SO and SO$_2$, can be easily
used by itself for estimating the age, because the ratios depend at
least as strongly
on the physical conditions and on the adopted grain mantle composition as
on the time.  Also, the abundance of atomic oxygen in the gas phase, if
not correctly accounted for, can seriously affect the chemistry.
The situation, however, is not totally hopeless, because
a careful comparison
between observations and model predictions can give some useful hints
on time estimates, and on the mantle composition.
Such a careful analysis
has to be done on each single source, however, for both the physical conditions
and mantle composition can vary from source to source,
so that the abundance ratios are not directly comparable.
In practice, a careful derivation of the molecular abundances
(which takes into account the source structure) coupled with a careful
modeling of the chemistry at the right gas temperature and density
is necessary.

  We applied our model to two well studied hot cores: Orion KL and IRAS16293. For the
S-bearing abundances towards Orion KL, we assumed that their emission arises
from the hot core region (which is strongly debatable) and  is not 
beam-diluted. We were not able to reproduce all of the observed 
abundances ratios with any of our models. The agreement with Model 2 
is satisfactory if we decrease the initial amout of atomic sulphur by 
a factor of 10. In that case, we derive a best age of $4\times 
10^3$~yr. However, the predicted abundance of CS is three orders of 
magnitude lower than the observed one.
Contrary to the case of Orion KL, the sulphur-bearing abundances
though the low mass hot core of IRAS16293-2422 have been carefully 
determined through a sophisticated model \citep{2002A&A...390.1001S},
which takes into account the density and temperature structure of the
source, as well as
the abundance profile of each studied molecule. 
 Using the standard value of cosmic ray rate, we found that Model
2, in which a large amount of atomic sulphur is initially present in 
the post-evaporative gas,
best reproduces the observed abundance ratios. In that case, we derived an age of $\sim
2\times 10^3$~yr from the evaporation era to the current stage of this particular low mass hot core.
If we decrease the initial amount of atomic sulphur in  Model 2 as 
for Orion KL, the agreement is still good and gives a similar age.
This analysis favors the hypothesis that sulphur is mainly evaporated from the
grains in the atomic form or in a form quickly converted into it. 
 On the contrary, if a higher rate is used as suggested by the recent modelling of \citet{2004doty}, best agreement occurs
with Model 1, where no atomic sulphur can be found in the grain 
mantle and only H$_2$S and OCS are initially present.
The strongest prediction of our atomic sulfur-rich model is the 
presence of large abundances of SO$_2$, derived from this form of 
sulfur, at late stages of hot cores. A futher systematic study of 
S-bearing-species towards older hot cores where the physical 
structure is well known would provide information to test this model. 
 Moreover, the fact that not all of the sulfur need be initially
in atomic form, given the reasonable agreement obtained using Model 
2', suggests that a signficant portion of the granular elemental 
sulphur may be tied up in materials such as iron sulphide 
\citep{2002Natur.417..148K}.

\begin{acknowledgements}

V. W. wishes to thank Franck Selsis
for helpful discussions on chemical modelling and uncertainties. V. 
W., C. C. and A. C. acknowledge support from PCMI.
P.C. acknowledges support from the MIUR grant "Dust and molecules in
astrophysical environments", and the ASI
grant (contract I/R/044/02).  E. H. acknowledges the support of the
  National Science Foundation (U. S.) for his research program in 
astrochemistry. The authors are grateful to Brunella Nisini and 
Malcolm Walmsley for useful discussions.
\end{acknowledgements}

\bibliographystyle{aa}

\bibliography{aamnem99,biblio}

\end{document}